  \documentstyle[prb,aps,multicol,epsfig]{revtex}
\begin{document}
\draft \date{\today} \title
  {Critical state in thin anisotropic superconductors of arbitrary shape}
\author{Grigorii P.~Mikitik}
  \address{B.~Verkin Institute for Low Temperature Physics \&
   Engineering, National Ukrainian Academy of Sciences,
   Kharkov 310164, Ukraine}
\author{Ernst Helmut Brandt}
  \address{Max-Planck-Institut f\"ur Metallforschung,
  D-70506 Stuttgart, Germany}
\maketitle

\begin{abstract}
 A thin flat superconductor of arbitrary shape and with arbitrary
 in-plane and out-of-plane anisotropy of flux-line pinning is
 considered, in an external magnetic field normal to its plane.
 It is shown
 that the general three-dimensional critical state problem for
 this superconductor reduces to the two-dimensional problem of
 an infinitely thin sample of the same shape but with a modified
 induction dependence of the critical sheet current. The methods
 of solving the latter problem are well-known. This finding thus
 enables one to study the critical states in realistic samples
 of high-$T_c$ superconductors with various types of anisotropic
 flux-line pinning. As examples, we investigate the critical
 states of long strips and rectangular platelets of high-$T_c$
 superconductors with pinning either by the $ab$ planes or
 by extended defects aligned with the $c$ axis.
\end{abstract}
    \begin{multicols}{2}   
    \narrowtext

\section{Introduction}  

   Numerous publications deal with the critical
 state problem in {\it isotropic} thin superconductors placed in
 a perpendicular magnetic field.\cite{1} By thin we mean that the
 sample thickness $d$ is considerably less than its lateral
 extension. Analytic solutions of the problem were obtained in
 the cases of a circular disk \cite{2}, a strip \cite{3}, and
 elliptic-shaped films.\cite{4}  For thin superconductors of an
 arbitrary shape, numerical methods for solving this problem were
 elaborated in Ref.\ \onlinecite{5,6}, and the critical state of a
 rectangular platelet was investigated in detail in
 Ref.\ \onlinecite{7}.  All the above-mentioned results can be
 derived considering the appropriate flat superconductors to be
 infinitely thin.  In the framework of this approach a possible
 {\it in-plane anisotropy} of pinning can be taken into
 account,\cite{4,8} yielding good agreement with appropriate
 magnetooptic experiments.\cite{8}

   It is important to note that in the highly anisotropic
 high-$T_c$ superconductors the flux-line pinning in general depends
 on the angle $\theta$ between the local direction of the magnetic
 induction ${\bf B}$ and the $c$ axis which in typical experiments
 is normal to the plane of the samples. For example, this
 type of anisotropy occurs when one takes into account the
 intrinsic pinning exerted by the CuO planes.\cite{9} Besides this,
 twin boundaries, columnar defects, and other extended defects give
 rise to such an out-of-plane anisotropy.\cite{1}  In all these cases,
 the flux-line curvature that always occurs in thin superconductors
 placed in a perpendicular magnetic field, leads to a dependence
 of the critical current density $j_c$ on the coordinate $z$ across
 the thickness of the sample. Therefore, the critical state
 problem becomes three-dimensional (3D),  and the feasibility of
 treating such superconductors as infinitely thin requires special
 consideration.

   For a thin anisotropic disk with radius $R$ placed in a
 perpendicular magnetic field it was shown \cite{10} that the small
 ratio $d/R$ enables one to  split the 2D critical state problem
 into two one-dimensional problems.
 The first one treats the critical state across the thickness of
 the disk, while the second one treats the disk as infinitely thin
 and gives the distributions of magnetic field and sheet current
 along its radius. Using this method, the magnetic moment of
 a thin anisotropic disk was theoretically studied \cite{11} in the
 case when the angular dependent $j_c$ has a peak at $\theta =0$.
 In this special geometry of the circular disk the directions of
 field and current are fixed and known in advance, and the
 critical state problem thus is  essentially not 3D but only 2D.
 This simplification does not take place in thin samples of
 arbitrary shape, where the current stream lines, magnetic
 contour lines, and penetrating flux fronts in general do not
 coincide.\cite{4} Moreover, according to Ref.\ \onlinecite{4},
 in the general case the directions of the circulating currents
 may change across the thickness of the sample, and a rotation or
 twist of the flux line arrangement is possible.

   In the present paper we show that even in the general 3D critical
 state problem of a thin superconductor with arbitrary shape and
 arbitrary anisotropy of $j_c$, the 3D pinning problem
 can be split into two simpler problems which can be solved:  The 1D
 problem across the thickness of the sample is solved analytically,
 while the in-plane problem reduces to that of the infinitely
 thin superconductor which was treated before. Since the methods of
 solving the latter problem are well elaborated,\cite{5,1} our
 finding enables one to study the critical state in flat samples of
 arbitrary shape and {\it with any type of anisotropy}.
 As concrete examples, we shall investigate the critical states of
 long strips and rectangular platelets with various types of
 out-of-plane anisotropy.

 \begin{figure}[F1] 
\epsfxsize= 0.80\hsize  \vskip 1.0\baselineskip
\centerline{ \epsffile{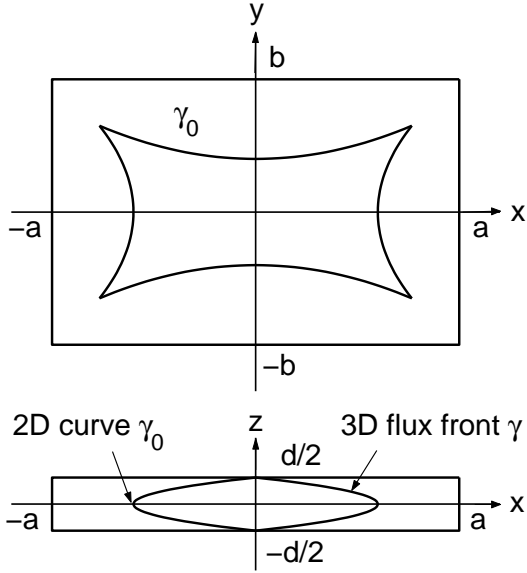}}
                        \vskip 0.5\baselineskip
\caption{The front of magnetic flux penetrating into a thin
 rectangular superconductor plate with pinning when the
 applied perpendicular magnetic field increases from zero.
 The top figure shows the two-dimensional curve $\gamma_0$
 that forms the equator of the three-dimensional flux front
 $\gamma$ shown in the lower plot. Inside this 3D front the
 magnetic induction ${\bf B}$ and current density ${\bf j}$ are
 exactly zero, and inside the curve $\gamma_0$ the perpendicular
 component $B_z$ is practically zero when $d\ll (a,b)$ for a
 thin plate with dimensions $2a \times 2b \times d$.
  } \end{figure} 

 \section{Splitting the 3D problem}  

   In what follows we assume that the pinning-caused characteristic
 magnetic field $H_{cr}= j_c d /2$ considerably exceeds the
 lower critical field $H_{c1}$, and the thickness $d$ of the sample
 is much greater than the London penetration depth  $\lambda$.
 Under these conditions the magnetic induction is practically equal
 to the magnetic field in the superconductor, and the so-called
 geometrical barrier \cite{12} is negligible. We study the critical
 state problem in the framework of the macroscopic approach, in which
 all considered lengths are larger than the flux-line spacing, and
 treat the superconductor as a uniform anisotropic medium.

   Let us place the coordinate system so that
 its $xy$ plane coincides with the middle plane of the sample
 and the $z$ axis is directed along the external magnetic field;
 the case of a rectangular platelet is shown in Fig.~1. Since
 the thickness $d$ of the sample is assumed to be much less than
 its smallest lateral dimension $L$, the strict equations
      $\nabla \cdot {\bf j} =0$
 and  $\nabla  \cdot {\bf B}=0$
 give that, in leading order in the parameter $d/L$, one has
 $j_z=0$, and the perpendicular field component $B_z$ is
 independent of $z$ inside the sample, i.e., $B_z= B_z(x,y)$.
 Accounting for the fact that the  scale of changes with
 $x$ and $y$  of $B_z$ and of the in-plane field
 ${\bf B}_t=(B_x,B_y)$ is $L$, while the appropriate scale for
 changes of ${\bf B}_t$ across the thickness is $d$, we may neglect
 the derivatives of ${\bf B}$ with respect to $x$ and $y$. The
 critical state equation can thus be written in the form:
      \begin{eqnarray}
      {\bf \hat z} \times  {\partial {\bf B}_t \over \partial z}
      = \mu_0\,{\bf j}_c \,,        \nonumber
      \end{eqnarray}
 where ${\bf \hat z}$ is the unit vector along $z$ and ${\bf j}_c$
 is the critical current density.
 This equation means that the critical gradient
 develops predominantly across the thickness of the sample, and we
 can consider the fields and the currents to be independent of $x,y$
 in sufficiently small parts of the sample with dimensions in
 the $x,y$ directions much greater than $d$ and much less than $L$.
 To proceed our analysis we write
       ${\bf B}_t=B_t{\bf \hat t}$ and ${\bf j}_c=j_c {\bf \hat n}$,
 where ${\bf \hat n}=(\cos\varphi,~\sin\varphi)$ and
       ${\bf \hat t} = (\cos\psi ,~\sin\psi)$.
 The angles $\varphi(x,y,z)$ and $\psi(x,y,z)$ define the directions
 of ${\bf j}_c$ and ${\bf B}_t$ in $xy$ planes. The above critical
 state equation can then be rewritten as
    \begin{eqnarray} 
    {\partial B_t \over \partial z} = \mu_0\, j_c
    \sin(\varphi -\psi) \,, \\
    B_t {\partial \psi \over \partial z} = -\mu_0\, j_c
    \cos(\varphi -\psi) \,.
    \end{eqnarray}
 We emphasize that in the general case the critical current density
 $j_c$ may depend on $B_z$, $B_t$, $\varphi -\psi$, $\psi$,  i.e.,
 $j_c =j_c (B_z, B_t, \varphi-\psi , \psi )$.  The
 quantities $B_t$ and $B_z$ can be also expressed in terms of the
 magnitude $B\equiv |{\bf B}|$ and the tilt angle $\theta$ between
 the local direction of ${\bf B}$ and the z axis:
      \[ B_z=B\cos \theta \,, ~~~
         B_t=B\sin \theta \,.  \]
 The dependence on $\psi$  occurs only if there is an anisotropy
 of pinning in the $xy$ plane (e.g., when twin boundaries exist
 in the sample\cite{13}). In the isotropic case, when the flux-line
 pinning depends neither on $\psi$ nor on the tilt angle $\theta$,
 the dependence of $j_c$ on $B_z$ and $B_t$ occurs only through the
 magnitude $B$.  Note also that,
 since $B_z$ is independent of $z$, it enters Eqs.~(1,2) as a
 parameter, and thus we have only {\it two} equations for the
 {\it three} functions $B_t$, $\varphi$, and $\psi$. One more
 equation can be obtained if one takes into account the prehistory
 of the critical state, telling how the flux lines were pushed
 into the sample.

    In increasing moderate magnetic field there is a flux and
  current-free core in the superconductor, Fig.~1. The surface of the
  core, $\gamma$, is the penetrating flux front, which we describe
  by a function $z=z_{\gamma}(x,y)$. In the limiting
  case of an infinitely thin superconductor ($d\to 0$) the flux
  front may be thought of as a flat curve $\gamma_0$, forming the
  outer rim (equator) of $\gamma$, since we have $B_z=0$ inside this
  contour. In the case of small but finite $d\ll L$, the component
  $B_z$ is still practically zero in the region inside $\gamma_0$
  since there
  the flux lines are almost parallel to the flat surfaces of the
  sample. Note that these flux lines may have different orientations
  at different values of $z$ if the sample is not a circular disk or
  a strip;\cite{4} thus, in principle, flux-cutting processes may
  occur in the superconductor.\cite{14,15,16}

     We now consider the third required equation in the region
  $z_\gamma(x,y) \le |z| \le d/2$ with $x,y$ inside the curve
  $\gamma_0$, i.e.\ between the lens-shaped empty core and the flat
  surfaces. In increasing applied field the flux lines are pushed
  continuously from the flat surfaces into this region, such that
  the in-plane orientation of the flux lines is a unique function
  of $B_t$, $x$, and $y$:
     \begin{eqnarray}  
     \psi =\Psi (B_t,x,y) \,.
     \end{eqnarray}
  The function $\Psi$ is found from the solution of the critical
  state problem for the infinitely thin sample (see below).
  Equation (3) is just the third required equation. Now we are able
  to solve Eqs.~(1--3). Dividing Eq.~(2) by Eq.~(1) we obtain
     \begin{eqnarray}  
     B_t {\partial \psi \over\partial B_t} =-\cot (\varphi -\psi)\,.
     \end{eqnarray}
  Inserting here the known field orientation, Eq.~(3), we find the
  difference $\varphi -\psi$ and the angle of the currents $\varphi$
  as functions of $B_t$. Finally, the dependence of $B_t$ on $z$ can
  be obtained from the implicit relation
     \begin{eqnarray}  
     z-z_{\gamma} =\int_0^{B_t}\!\!\! {db_t \over
     \mu_0 j_c(0,b_t,\pi /2,\Psi (b_t,x,y))} \,,
     \end{eqnarray}
  which follows from Eq.~(1), and from the formula
    \begin{eqnarray}
    j_c (0, B_t, \varphi-\psi , \psi )=
    {j_c (0, B_t, \pi /2, \psi ) \over \sin (\varphi -\psi)} \,,
    \nonumber
    \end{eqnarray}
  reflecting the fact that only the component of $j_c$ normal to
  ${\bf B}$ is essential (in absence of flux cutting). The function
  $z_{\gamma }(x,y)$ describing the shape of the flux front is
  obtained from the equality:
     \begin{eqnarray}  
     {d \over 2} -z_{\gamma} = \int_0^{B_s}\!\!\!
     {db_t \over \mu_0 j_c(0,b_t,\pi /2,\Psi(b_t,x,y) )} \,,
     \end{eqnarray}
  where $B_s(x,y) = B_t(x,y,d/2)$ is the tangential component of the
  surface field. This $B_s$  may be obtained by solving the critical
  state equation for infinitely thin superconductors. Thus,
  Eqs.~(3--6) give the solution of the critical state problem across
  the thickness of the sample in the region inside the curve $\gamma_0$.

    It should be emphasized that the above formula
  for the dependence of $j_c$ on $\varphi -\psi $ and Eqs.~(3,5,6)
  hold only if flux cutting does not occur in the superconductor when
  the critical state is established in it. Flux cutting will
  only happen if $|\cot (\varphi -\psi )|$, calculated from Eq.~(4),
  turns out to be greater than some critical value $\chi$.\cite{15}
  In this case the so-called generalized critical-state model \cite{15}
  must be used to describe the critical state in the region of the
  sample inside $\gamma_0$. However, since at present we cannot
  point out the shape of the superconductor for which this
  flux cutting really occurs, we do not analyze this case here.

     Let us now derive the third equation for the region where
  $B_z\neq 0$, i.e., outside the curve $\gamma_0$. In deriving
  it is necessary to take into account that in an
  anisotropic superconductor the direction of the flux-line velocity
  in general does not coincide with the direction of the
  Lorentz force causing this movement, and thus the appropriate
  electric field ${\bf E}$ is not always parallel to the current
  density (see Appendix A). With this in mind, the third
  equation results from the following two conditions:
  First, the shape of a flux line cannot change when an applied
  magnetic field $H$ increases slightly so that $\gamma_0$ and
  all the flux lines in this region of the sample are shifted by
  distances which are much less than the sample width $L$.
  Second, during the shift, each line element moves along the local
  direction of $[{\bf E}\times {\bf B}]$. The geometrical analysis
  of these conditions gives
     \[
      {B_z^2\sin \xi -B_t^2\cos \psi \sin(\psi-\xi) \over
      B_z^2\cos \xi +B_t^2\sin \psi \sin(\psi-\xi)}={\rm const}
     \]
 where the angle $\xi$ is determined by Eqs.~(A2,A3,A5)
 in Appendix A. Since $B_t(0)=0$, the obtained relation can be
 rewritten as
     \begin{eqnarray}  
     \tan [\xi - \xi (0)]={B_t^2 \sin [\psi - \xi (0)]
     \cos [\psi -\xi (0)] \over
     B_t^2 \cos^2 [\psi -\xi (0)] + B_z^2}\ ,
     \end{eqnarray}
 with $\xi (0)\equiv \xi (x,\,y,\,0)$.
 This is the third required equation in the region
 where $B_z\neq 0$. When an in-plane anisotropy of the flux line
 pinning is absent, it is possible to find the solution of
 Eqs.~(1,2,7) in the interval $0 \leq z \leq d/2$.
 The initial conditions to these equations are
     \begin{eqnarray}  
     B_t(0)=0 \,, ~~~ \varphi (0) -\psi(0)=\pi /2 \,,
     \end{eqnarray}
 where the last equality follows from Eq.~(2) with $B_t(0)=0$.
 Since the solution of the equations is unique, it is sufficient
 to guess it. The solution has the form
     \begin{eqnarray}  
     \varphi (x,y,z) = \varphi_0 (x,y) \,, \\
     \psi (x,y,z)=\varphi_0 (x,y)-\pi /2 \equiv \psi_0(x,y) \,,\\
     z = \int_0^{B_t}\!\!\! {db_t \over
     \mu_0 j_c(B_z,b_t,\pi /2,\psi_0 )} \,.
     \end{eqnarray}
 Here the angle $\varphi_0(x,y)$ defines the direction of the sheet
 current in the critical state of the infinitely thin superconductor
 and is assumed to be known. In obtaining the solution we have taken
 into account that $\xi =\varphi$ at $\varphi -\psi =\pi/2$ when the
 in-plane anisotropy is absent.
 Eqs.~(9--11) mean that the flux lines are {\it curved} in the region
 where $B_z\neq 0$, but they are {\it not twisted} there. In other
 words, the distributions of the magnetic induction and the current
 across the thickness of the sample are the same as in a circular
 disk or a strip if one considers small regions in the $xy$ plane
 with the same values of $B_z$. Eq.~(11) allows us to find an
 implicit relation between the sheet current $J$ and $B_z$:
     \begin{eqnarray}  
     {d \over 2} = \int_0^{J/2} \!\!\! {d h_t \over
     j_c(B_z, \mu_0 h_t,\pi /2,\psi_0 )} \,.
     \end{eqnarray}
 Here we have used that $B_t(z=d/2) \equiv B_s= \mu_0 J/2$, where
 $J\equiv |{\bf J}|$ and ${\bf J}$ is the current density integrated
 over the film thickness.
 It should be noted that
 although we indicate explicitly the angle $\psi_0$ as an argument
 of $j_c$ in Eqs.~(11,12), the critical current density is
 independent of this angle in cases without in-plane anisotropy.

   We now discuss the case when some in-plane anisotropy of
 flux-line pinning exists in the superconductor together with
 the out-of-plane anisotropy, i.e.\ we allow for a dependence of
 $j_c$ on $\psi$. This case can occur, e.g.\ in
 twinned crystals of high-$T_c$ superconductors.
 When an in-plane anisotropy exists, it is impossible
 to find the solution of Eqs.~(1,2,7) in the general form.
 We may only state that in this situation the in-plane angle $\psi$
 generally depends on $z$, and thus flux lines are {\it not only
 curved but also twisted}. Then, according to Eq.~(2), the current
 is not normal to the magnetic induction. The magnitude of the
 twist is determined by the relative strengths of the in-plane
 and out-of-plane anisotropies (and, of course, by
 the thickness of the sample).
 Interestingly, when $j_c$ does not depend on $B_t$, the solution
 described by Eqs.~(9--11) becomes true again, and the twist
 disappears. (This situation occurs, e.g., in sufficiently large
 magnetic fields, $H\gg H_{cr}$, or if there is no
 out-of-plane anisotropy in the superconductor and $j_c$ is
 practically independent of the magnitude of the magnetic
 induction). However, if $H$ changes, in this case a flux line is
 shifted in a direction which does {\it not} lie in the plane
 containing the line. According to the results of Appendix A,
 the direction makes an angle $\alpha$ with the above-mentioned
 plane where
   \[
   \tan\alpha=-{\partial [\ln j_c(B_z, \varphi-\psi_0,
   \psi_0)] \over \partial \varphi}\Big|_ {\displaystyle{\varphi=\psi_0+}
   {\pi\over 2} } \ .
   \]

 To sum up,
 it is important to emphasize that at a given dependence of $j_c$
 on $B_z$, $B_t$, $\varphi-\psi$ and $\psi$, the one-dimensional
 equations (1), (2), (7) with initial conditions (8) can be solved
 (at least numerically) in the case of an {\it arbitrary} anisotropy
 of flux-line pinning, and a relation $J(B_z, \varphi_0)$
 generalizing Eq.~(12) can be obtained. Here $\varphi_0$ defines
 the direction of the sheet current in the $xy$ plane.

   Thus, we have reduced the 3D critical state problem to a
 2D problem in which the $B_z$-dependence of the sheet
 current $J(B_z, \varphi_0)$ is determined not only by
 the $B_z$-dependence of the critical current density $j_c$
 but also by its dependence on the tilt angle
 $\theta$ of the flux lines away from the film normal, see Eq.~(12).
 It should be noted, however, that to obtain this reduction we have
 assumed that $d$ is much less than the characteristic scale of changes
 of ${\bf B}$ in the $xy$ plane. Numerical solutions of the
 critical state equation for infinitely thin superconductors
 of various shapes show that this assumption indeed is true
 everywhere; the scale of variation is of the order of $L$ except
 for a small region near the flux front $\gamma_0$ where
 the derivatives $\partial B_x/\partial y$ and
 $\partial B_y/\partial x$  are not small compared with
 $\partial {\bf B}_t/\partial z$. However, the width of this region
 is determined by the thickness $d$ and is thus small.

   Knowing the function $J(B_z,\varphi_0)$ obtained from Eq.~(12)
 or from the above-mentioned generalization of this equation,
 it is possible to solve the critical state equations for an
 infinitely thin superconductor of any shape.\cite{1,5,6}  The
 {\it static} equations to be solved are
    \begin{equation}
    \nabla \cdot {\bf J}=0
    \end{equation}
 and the Biot-Savart law connecting $B_z$ with the sheet current
 ${\bf J}(x,y) = \int_{-d/2}^{d/2} {\bf j}(x,y,z)\, dz$,
     \begin{equation}
    \mu_0^{-1}B_z({\bf r})=H + \int_S
    {[{\bf R}\times {\bf J}] \over 4\pi R^3} d^2 r' \,,
    \end{equation}
 where ${\bf R}={\bf r}-{\bf r'}$; ${\bf r}$ and ${\bf r'}$ are
 2D vectors in the $xy$ plane, and the integration is carried out
 over the specimen area $S$. In addition, one has
 to take into account that $B_z=0$ inside the 2D flux front
 $\gamma_0$, while outside $\gamma_0$ the sheet current is equal to
 $J(B_z,\varphi_0)$. At given applied field $H$, a solution of this
 static problem exists only for a certain curve $\gamma_0$, which
 determines the shape and position of the penetrating flux front.
 In Refs.~\onlinecite{1,5,7}, in fact, the more general
 {\it dynamic} equations were considered, which allow one to find
 the solution of the 2D critical state equations very effectively
 for any prehistory $H(t)$.
 The appropriate formulas are presented in Appendix B.

   Solving these equations for the infinitely thin superconductor
 of arbitrary shape in a perpendicular applied field $H$, one
 obtains $B_z(x,y,H)$, $J(x,y,H)$, $\varphi_0 (x,y,H)$,
 $B_s(x,y,H) = \mu_0 J/2$, the direction of ${\bf B}_s$,
 i.e., the function $\psi_0(x,y,H)=\varphi_0 (x,y,H)-\pi /2$,
 and also the position of $\gamma_0$. Eliminating $H$ from
 $\psi_0(x,y,H)$ and $B_s(x,y,H)$ in the region inside $\gamma_0$,
 and replacing $B_s$ by $B_t$ we arrive at the function
 $\Psi (B_t,x,y)$ that enters Eq.~(3). Thus, the solution of the
 2D critical state problem for the infinitely thin superconductor
 enables one to obtain all the necessary information for the
 calculations of the core shape $z_{\gamma}(x,y)$ and
 distributions of $B_t$, $\psi$, $\varphi$ across $z$ at any $x$
 and $y$. The knowledge of all these functions permits to describe
 the 3D critical state of thin anisotropic superconductors in detail.

 \section{Out-of-plane Anisotropy}

   We now use the results obtained above to investigate the critical
 states of thin superconductors with out-of-plane anisotropy alone.
 That is, we consider the case when the critical current density
 $j_c(B_z,B_t,\pi /2,\psi )$ does not depend on the in-plane angle
 $\psi$. In addition we here neglect its dependence on $B$ such that
 $j_c$  depends only  on the flux-line tilt angle $\theta$,
 $j_c =j_c(\theta )$. This situation occurs when the
 flux-line pinning is isotropic in the $xy$ plane, and the scale
 $B_0$ characterizing the dependence of $j_c$ on $B$ considerably
 exceeds the  self-fields of the critical currents. As shown above,
 the anisotropy of $j_c(\theta )$ can be accounted for
 by introducing an appropriate $B_z$ dependence of the
 sheet current  $J=J(B_z)$, implicitly given by
  \linebreak

  \begin{figure}[F2]  
\epsfxsize= 0.75\hsize  \vskip 1.0\baselineskip
\centerline{ \epsffile{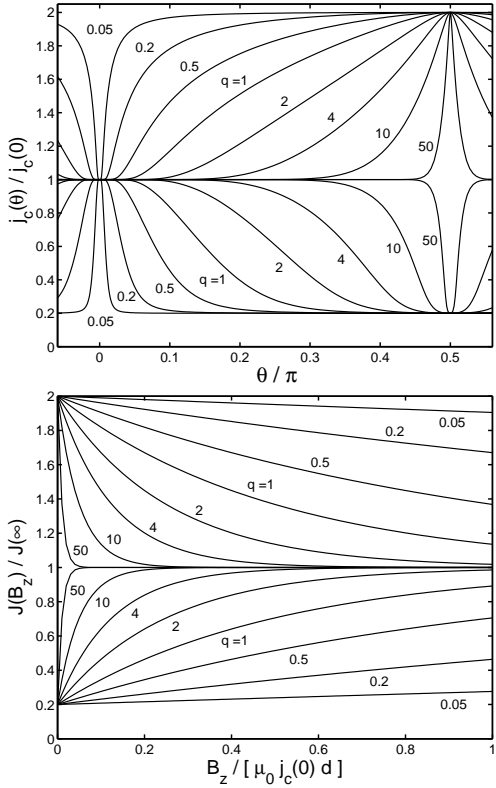}}
                       \vskip 0.5\baselineskip
\caption{The out-of-plane anisotropy of the critical current
 density $j_c(\theta)$ described by the model (18) (top plot)
 and the equivalent induction dependence of the sheet current
 $J(B_z)$ of Eq.~(17) (lower plot). In both plots the upper half
 is for the parameter $p=1$ and the lower half for $p= -0.8$.
 The second parameter is $q=0.05 \dots 50$.
  } \end{figure}     

  \noindent
 Eq.~(12). In the considered case, a more explicit form of
 this dependence may be
 obtained in the following way. Differentiation of  Eq.~(12)
 with respect to $B_z$  yields the two equations:
     \begin{eqnarray} 
     j_c(\theta) d = J(B_z) -B_z{d J(B_z) \over d B_z} \,, \\
     \tan \theta   = { \mu_0 J(B_z) \over 2 B_z } \,.
     \end{eqnarray}
 In fact, Eqs.~(15,16) are the parametric form of that function
 $j_c(\theta )$ which results in a given dependence $J(B_z)$
 via Eq.~(12).

    To fix ideas, we consider the following rather
 general model dependence for $J(B_z)$:
    \begin{eqnarray}  
    J(B_z) = j_c(0) d \left [ 1 + p\, \exp\left
    (-{q \,B_z \over B_{cr}}\right ) \right ] \,,
    \end{eqnarray}
 where $B_{cr} = \mu_0 j_c(0) d /2$, while
 $j_c(0)$, $p$ and $q$ are the parameters of this example
 ($p$ and $q$ are dimensionless). One then easily
 verifies that the corresponding angular dependence of the
 critical current density takes the form:
     \begin{eqnarray}  
     j_c(\theta) =j_c(0)\left [ 1 + p \, (1+ q\, t)
     \exp(- q\, t)\right ] \,, \nonumber \\
     \tan \theta = t^{-1}\left [ 1 + p\,
     \exp(- q\, t)\right ] \,,
     \end{eqnarray}
 where $t$ is a curve parameter with range $0 \le t \le \infty$
 equivalent to $\pi/2 \ge \theta \ge 0$. This model dependence
 $j_c(\theta)$ is presented in Fig.~2 together with the
 corresponding $J(B_z)$, Eq.~(17). For appropriate choices of
 the parameters $p$ and $q$ this model describes the intrinsic
 pinning by the CuO planes
 in high-$T_c$ superconductors ($p>0$, maximum $j_c$ at
 $\theta=\pi/2$) and pinning by columnar defects
 perpendicular to the film ($p<0$, maximum $j_c$ at $\theta=0$).
 It may thus be used to simulate these important cases. We remark
 that the critical state of a circular disk with some model for
 intrinsic pinning was studied numerically in Ref.~\onlinecite{17},
 while the magnetization of the disk in the case when $j_c$ has
 a peak at $\theta=0$ was analyzed in Ref.~\onlinecite{11}.
 However, in those papers only the fully penetrated critical state
 was considered.

   Below we calculate the partially and fully penetrated critical
 states in strips and rectangular platelets with anisotropic
 pinning using the angular dependence $j_c( \theta )$ described
 by Eqs.~(18).  The corresponding results for circular disks are
 very similar to the results for the strip.
 In fact the algorithm \cite{18} used below to time-integrate
 the 1D equation of motion for the sheet current $J(x,t)$ in thin
 strips is almost identical to the algorithm \cite{19} used
 for thin disks.

 \section{Thin Strip}

    We begin with the calculation of the critical state in an
 infinitely thin strip (see Appendix B). In this calculation the function
 $J(B_z)$ described by Eq.~(17) is considered as the critical
 value of the sheet current. The obtained results are presented
 in Figs.~3--7.

    Figure 3 shows the induction profiles $B_z(x)$ and sheet
 current profiles $J(x)$ in increasing $H$ for thin strips with
 various anisotropies  of $j_c(\theta)$ depending on
 the parameters $p$ and $q$.  The profiles are presented for
 several values of $x_0$, the position of the 2D flux front
 $\gamma_0$. The reference case of isotropic pinning ($p=0$) is
 also shown here. In Fig.~3a
 one can see that for $p>0$ (pinning by the CuO layers)
 the penetrating flux-density profile is very steep at $x=x_0$:
 $B_z(x)$ jumps almost abruptly and $J(x)$ has a very sharp
 peak which ideally should reach the maximum value
 $J(x_0) = (1+p) j_c(0) d$, cf.\ Eq.~(17).
 In our computation this ideal maximum height is not reached
 due to  our finite creep exponent $n=101$  and finite number
 $N=300$ of equidistant grid points (to illustrate the influence
 of $n$ the case of a small value $n=11$ is shown in Fig.~3d).
 As opposed to this, in Figs.~3b and 3c
 the induction profiles  for $p<0$  (pinning by linear
 defects perpendicular to the strip plane) are less sharp:
 at $x=x_0$, $B_z(x)$ vanishes less steeply (almost linearly
 for Fig.~3b)  and $J(x)$ decreases
 monotonically and exhibits an inflection point with
 infinite slope, $dJ/dx|_{x=x_0} = \infty$,
 $J(x_0) = (1+p)j_c(0) d$. These
 general features of the  flux front of superconductors with
 \linebreak

\epsfxsize= 0.79\hsize  \vskip .0\baselineskip
\centerline{ \epsffile{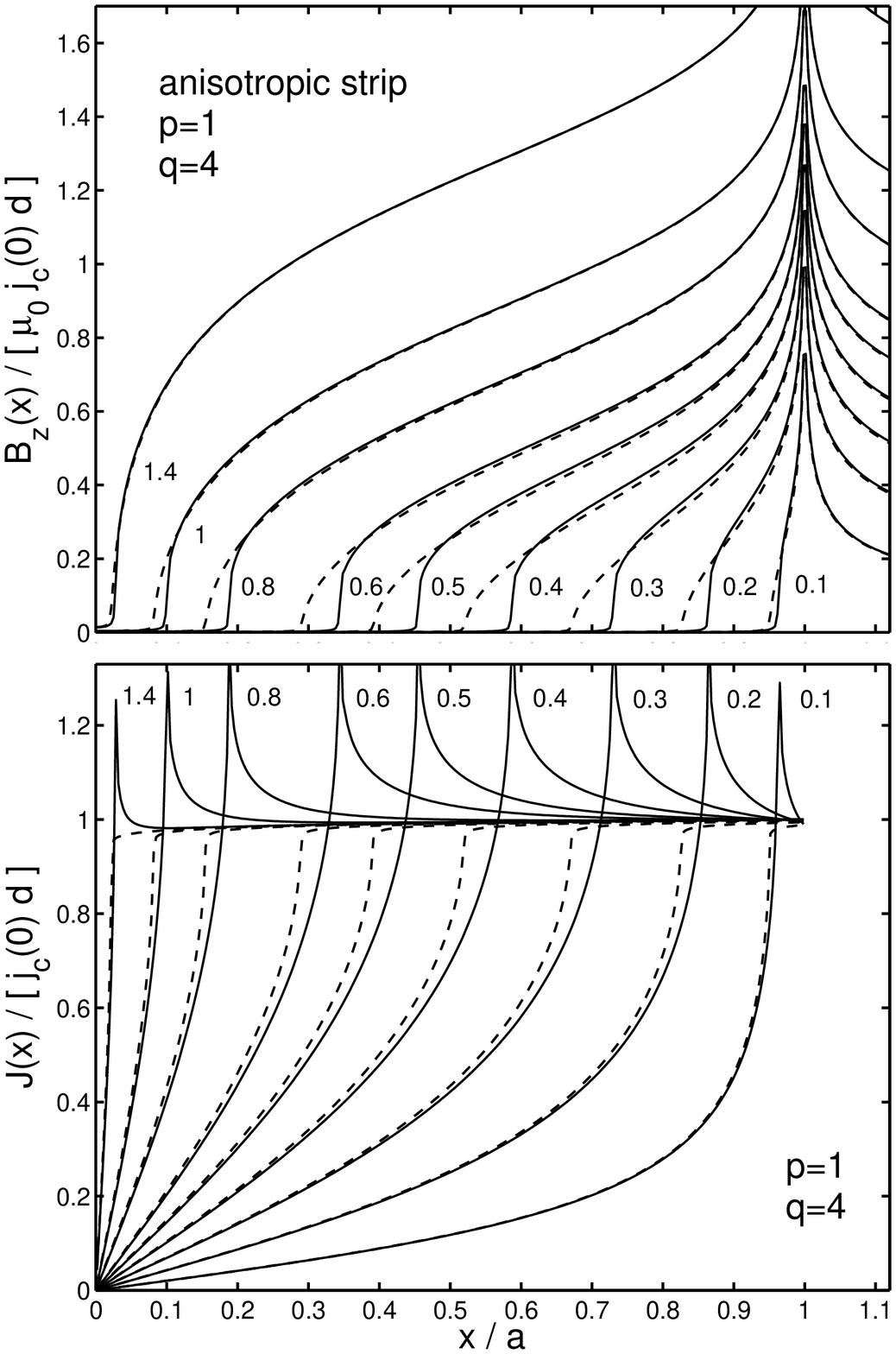}}
    \vspace*{-1cm} \hspace*{.2cm} (a) \vspace{.7cm}
\epsfxsize= 0.79\hsize  \vskip .5\baselineskip
\centerline{ \epsffile{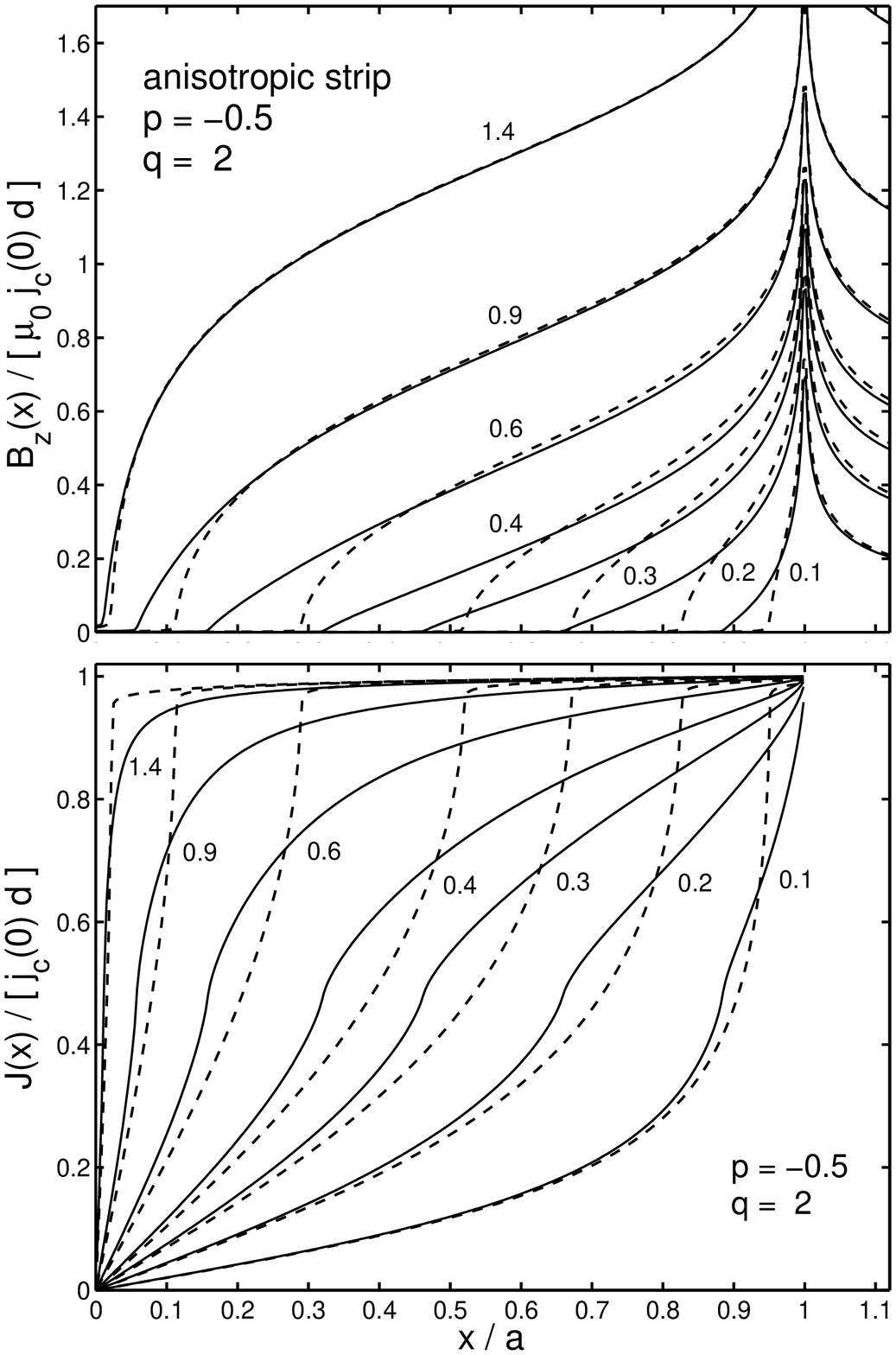}}
    \vspace*{-1cm} \hspace*{.2cm} (b) \vspace{1cm}
\epsfxsize= 0.79\hsize  \vskip .0\baselineskip
\centerline{ \epsffile{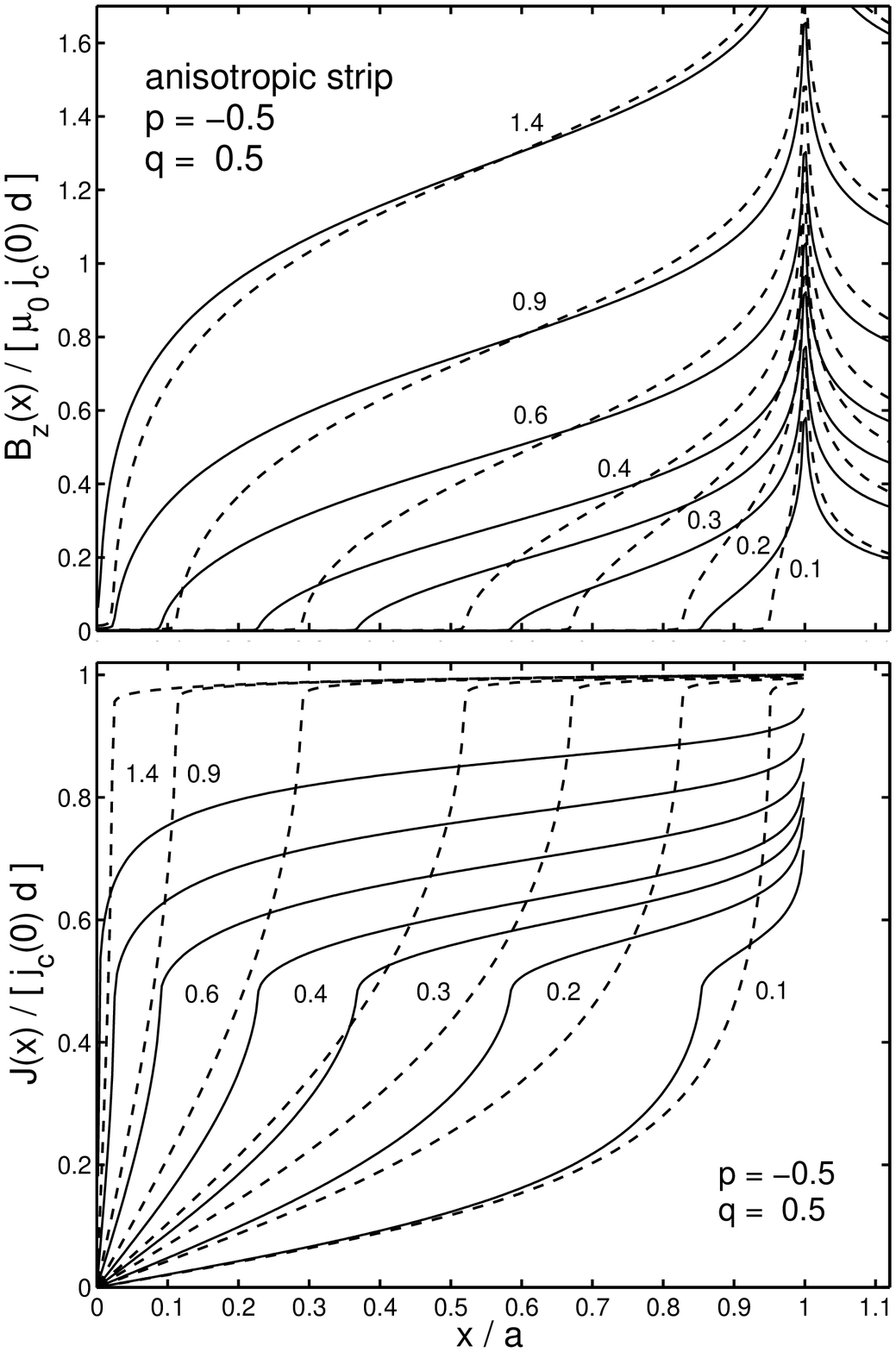}}
     \vspace*{-1cm} \hspace*{0.2cm} (c) \vspace{.7cm}
\epsfxsize= 0.79\hsize  \vskip .5\baselineskip
\centerline{ \epsffile{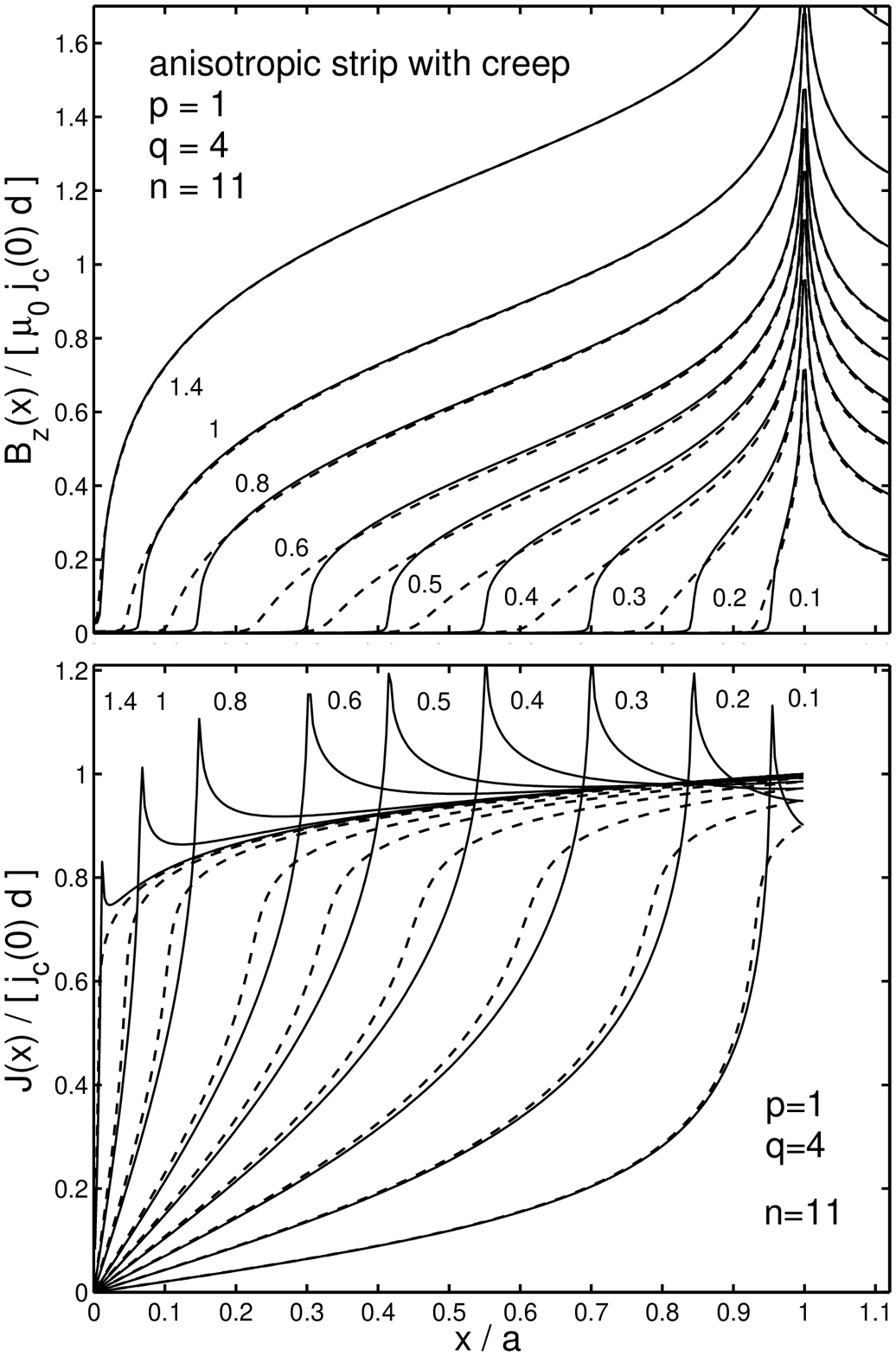}}
     \vspace*{-1cm} \hspace*{0.2cm} (d) \vspace{1cm}
 \end{multicols}
 \widetext
 \begin{figure}[F3]
                         \vskip .0 \baselineskip
\caption{Spatial profiles of the perpendicular field component
 $B_z(x)$ (upper plot) and of the sheet current $J(x)$
 (lower plot) of a thin strip with anisotropic pinning
 described by the model (18), see  Fig.~2. The
 various curves correspond to increasing applied field
 $H=0.1,\, 0.2,$  $\dots$ 0.8, 1, 1.4 in units of $j_c(0)d$.
 The dashed curves are for isotropic pinning ($p=0$) and the
 solid curves for anisotropic pinning with parameters:
 a) $p=1$, $q=4$. b) $p=-0.5$, $q=2$. c) $p=-0.5$, $q=0.5$.
 Creep exponent $n=101$.
 To check the influence of $n$,  the same case
 as in a) is shown in d) for $n=11$.
} \end{figure}  
  \newpage
  \begin{multicols}{2}   
  \narrowtext

 \begin{figure}[F4]  
\epsfxsize= 0.95\hsize  \vskip .5\baselineskip
\centerline{ \epsffile{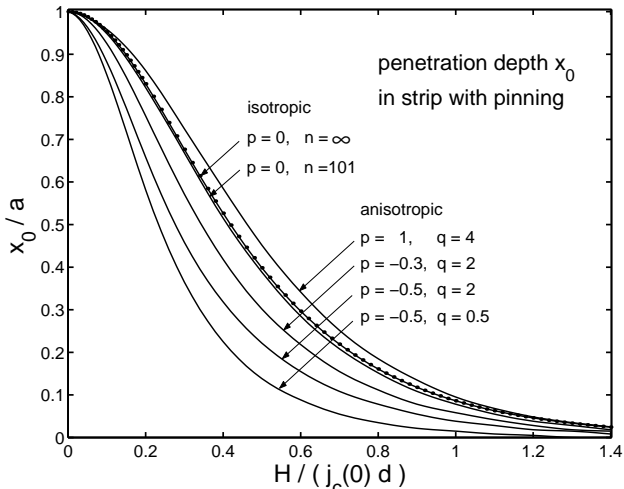}}
                        \vskip 0\baselineskip
\caption{The $x$ coordinate $x_0(H)$ of the penetrating flux
 front, or the penetration width $a-x_0$, for thin strips with
 various anisotropies of pinning $j_c(\theta)$, Eqs.~(18),
 calculated with large creep exponent $n=101$. The line marked
 by dots shows the static analytical result for isotropic
 strips, Eq.\ (20).
  } \end{figure}  

 \noindent
 anisotropic pinning are
 derived analytically in a forthcoming paper.\cite{20}
 It turns out that in the vicinity of $x_0$ the profiles are
 $B_z(x) \propto (x-x_0)^{\beta}$ and
 $J(x)-J(x_0) \propto |x-x_0|^{\beta}$ where
 $\beta \approx 0.5 - \pi^{-1}\arctan(pq)$.

   In the isotropic case ($p=0$) our numerical results practically
 completely coincide with the exact solution\cite{3}
   \begin{eqnarray} 
   {J(x) \over j_c d} = {2\over \pi} {\rm arccot} \Big\{
   \Big[ {\rm max}\Big( 0,\, {(x_0^2-x^2)a^2 \over (a^2-x_0^2)x^2}
   \Big) \Big]^{1/2} \Big\} \,, \\
   x_0= {a \over \cosh(\pi H /j_c d) } \,.
   \end{eqnarray}
 where $0\le x\le a$ and $2a$ is the width of the strip.
 It should be also mentioned that in the case
 of monotonically decreasing $J_c(B_z)$ the critical states of
 a thin strip and a thin disk were studied in
 Refs.~\onlinecite{21,22}, respectively. In Ref.~\onlinecite{21}
 the Kim model $J_c(B_z)=J_c(0)B_0/(B_0+|B_z|)$
 was considered, while in Ref.~\onlinecite{22} the Kim model and the
 exponential
 dependence $J_c(B_z)=J_c(0)\exp(-|B_z|/B_0)$ were analyzed
 where $J_c(0)$ and $B_0$ are some constant parameters. Our results
 in the case $p>0$ are similar to the results in these papers.
 Equations~(15,16) show that both models\cite{21,22} may be
 interpreted as angular dependences of $j_c(\theta)$ with maxima
 at $\theta=\pi/2$ if $B_0$ is of the order of $J_c(0)$.

   The position $x=x_0$ of the flux front $\gamma_0$ as a function
 of the increasing applied field $H$ is shown in Fig.~4
 for the same anisotropic strips. One realizes that for $p>0$
 the penetration depth $a-x_0$ is smaller, and for $p<0$ larger,
 than for the isotropic case $p=0$. The isotropic case is shown
 twice in Fig.~4: The numerical result for $x_0$ obtained with
 a finite creep exponent $n=101$ is only slightly smaller than the
 ideal static result, Eq.~(20). This demonstrates the accuracy of
 our computations.

   Using the above results for the infinitely thin strip and
 equations of Sec.~II, we can now describe the {\it two dimensional}
 critical state of the {\it anisotropic} strip with small but
 {\it finite} thickness.

 \begin{figure}[F5]  
\epsfxsize= 0.95\hsize  \vskip 0\baselineskip
\centerline{ \epsffile{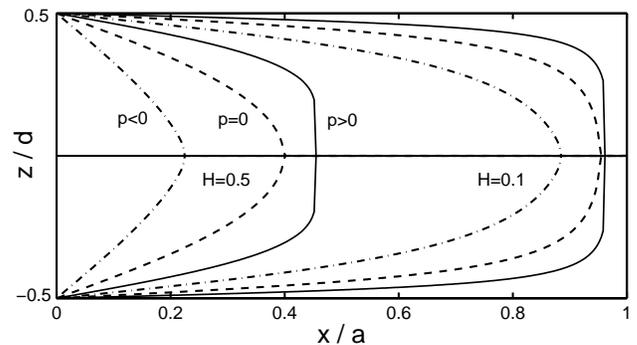}}
                        \vskip 0.5\baselineskip
\caption{Flux fronts $z_\gamma(x)$ in thin strips in increasing
 field $H$ calculated from Eq.~(21) and the data $J(x)$
 for isotropic pining ($p=0$, dashed lines) and  anisotropic
 pinning with $p=1$, $q=4$ (solid lines)
 and $p=-0.5$, $2$ (dash-dotted lines). Shown are two applied
 fields $H= 0.1$,  $0.5$ in units of $j_c(0) d$. Note the
 different scales for $x$ and $z$.
  } \end{figure}   


   The general 3D flux front $\gamma$ (Fig.\ 1) is determined by
 Eq.~(6) which is rewritten for our model Eqs.~(17,18) as
   \begin{eqnarray} 
   z_\gamma(x,y) ={d\over 2}\Big[ 1 -{J(x,y)\over J_{\gamma_0}}
       \Big] \,,
   \end{eqnarray}
 where $J_{\gamma_0} =(1+p) j_c(0) d$ and $J(x,y)$ is the value of
 the sheet current at a given point inside $\gamma_0$. Fig.~5 shows
 $z_{\gamma}(x)$, the spatial profile or cross section of flux
 fronts $\gamma$, in thin strips with anisotropic pinning.
 Note that for $p>0$ the
 fronts have a wide flat part near the central plane $z=0$, while
 for $p<0$ the fronts have a more rounded shape near $x=x_0$ as
 compared to  the isotropic case $p=0$.

   Physically, the wide flat flux front for $p>0$ results
 from the fact that in this case the flux lines oriented
 perpendicularly to the film (and to the CuO planes) are less pinned
 and can thus penetrate more easily than the flux lines
 oriented parallel to the film plane. In the opposite case $p<0$,
 flux lines in the film plane move more easily and cause a more
 pointed wedge-like front.

   The critical states of thick {\it isotropic} strips and disks
 were calculated in Refs.~\onlinecite{23,24} using the appropriate
 2D  equations. The isotropic fronts ($p=0$)
 obtained from Eq.~(21) and depicted in Fig.~5 coincide
 with the flux fronts computed for such strips \cite{23} and
 disks \cite{24} if one takes $d \ll a$ and $j_c=j_c(0)$.
 These flux front profiles are equal for strips and disks since
 the appropriate static sheet currents
 $J(x)$ and $J(r)$ have identical forms \cite{2,3} described by
 Eqs.~(19,20); for the disk $x$ should be replaced by the
 radial coordinate $r$ and $a$ means the disk radius.
 Interestingly, the thin film solution can well describe features
 of thick strips and disks up to quite large aspect ratios
 $d/2a \approx 0.2$ as long as the flux front is not too close
 to the center ($x_0 > d$).

 \begin{figure}[F6]  
\epsfxsize= 0.90\hsize  \vskip 0\baselineskip
\centerline{ \epsffile{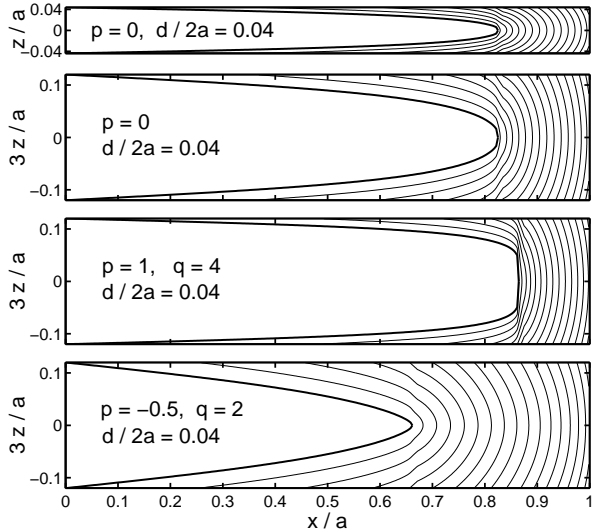}}
                        \vskip 0.5\baselineskip
 \caption{Magnetic field lines in thin strips with aspect ratio
$d/2a = 1/25$ in a perpendicular magnetic field $H =0.2 j_c(0)d$.
Isotropic pinning ($p=0$, top) and same anisotropic pinning
examples as in Figs.\ 3--5 are shown. The upper plot is to scale,
the lower three plots are stretched along $z$ for clarity.
The depicted lines are contours of the vector potential with
nonequidistant levels $A_i \propto i^2$, $i=0,1,2,3,\dots$, and are
thus directed along the magnetic field. The nearly equidistant lines
indicate a nearly constant gradient of the field magnitude. The
inner contour $A(x,z)=0$ coincides with the flux front $\gamma$.
   } \end{figure}  

    In a strip the sheet current has a fixed direction, and thus
 there is no rotation of the flux lines in the region inside
 $\gamma_0$, i.e., $\psi=$const in Eq.~(3). On the other hand,
 since in the considered case any in-plane anisotropy of pinning
 is absent, the flux lines are not twisted in the region
 $x\ge x_0$, and the solution (9--11) is valid.
 The distribution of $B_t=B_x(x,y,z)$
 in the strip can be found using Eqs.~(5,11). For our model
 (17,18), these equations are solved analytically, yielding
 for $x \le x_0$
     \begin{eqnarray}  
     B_x(x,z) \,=0 ,\,~~~~~~~~~~~~~~ |z|\le z_{\gamma}(x) \,,
      \nonumber \\ \nonumber
     |B_x(x,z)|= j_c(0)(1+p)\ [|z|-z_{\gamma}(x)] , \\
     ~~~z_{\gamma}(x)\le |z| \le d/2 \,,
     \end{eqnarray}
 while in the region $x\ge x_0$ we arrive at
    \begin{eqnarray}  
     |B_x(x,z)| = \nonumber
     j_c(0) |z|\! \left [ 1 + p\, \exp\left( \!
     -{q \,B_z(x)d \over 2|z|B_{cr}}\right) \right ] , \\
        ~~~ |z|\le d/2 \,.
    \end{eqnarray}
 Note that in the anisotropic case ($p\ne 0$) $B_x$ is not
 linear in $z$ in the region $x\ge x_0$. Taking into account these
 equations and the fact that $B_z$ is
 independent of $z$, one finds the 2D distributions of ${\bf B}$
 inside the thin anisotropic strips shown in Fig.~6.
 As can be seen, in the isotropic case our approach gives the
 expected\cite{23} magnetic field lines except in a narrow region near
 the flux front, where this approximation breaks down (see Sec.~II).
 The width of this region is less than, or of the order of $d$.

 \begin{figure}[F7]  
\epsfxsize= 0.90\hsize  \vskip 1.5\baselineskip
\centerline{ \epsffile{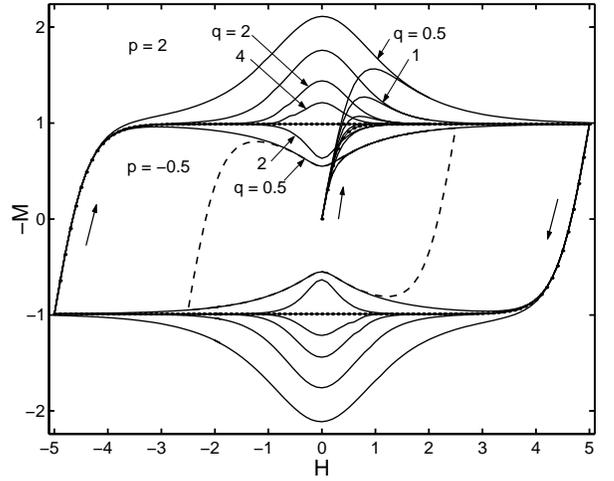}}
                       \vskip 0.5\baselineskip
 \caption{Magnetization curves of thin strips with anisotropic
 pinning, Eqs.~(17,18). The dotted line is for isotropic pinning
 ($p=0$). The loops with a hump are for $p=2$ and the loops with
 a dip for $p=-0.5$. The dashed curve has half the sweep amplitude.
 The applied field $H$ is in units of $j_c(0) d$ and $M$ is the
 magnetic moment per unit length in units of $j_c(0)d a^2$.
 Creep exponent $n=51$.
   } \end{figure}  

    Knowing the direction of ${\bf B}(x,z)$ we can further
 calculate the distribution of $j=j_c$ in the region outside
 the 3D front  $\gamma$ (inside $\gamma$ one has $j=0$).
 When $x < x_0$, the
 flux lines are practically parallel to the $xy$ plane, and thus
 $j_c=j_c(\pi/2)=j_c(0)(1+p)$. If $x>x_0$, $j_c$ is given
 by the first equation (18) with $t$ replaced by the ratio
 $B_z(x)/[\mu_0j_c(0)z]$. Note
 that for $p>0$ and $x>x_0$, $j_c(x,z)$ is greater near the surfaces
 ($|z|\approx d/2$) than in the central plane of the strip ($z=0$).
 The opposite situation occurs when $p<0$.

    In Fig.~7 the virgin magnetization curves and magnetic hysteresis
 loops $M(H)$ are presented for strips with various anisotropies
 of pinning. For $ab$ plane pinning ($p>0$) a so-called central peak
 appears. Thus, in high-$T_c$ superconductors this often observed
 peak \cite{25} in principle may be caused by intrinsic pinning by the
 CuO planes.\cite{17}  Note that in this case the width of the peak is
 proportional to the thickness $d$ of the sample (our field units
 are $j_c(0) d$). From this one may estimate the contribution
 of the intrinsic pinning when the peak is observed.
 In the opposite case $p<0$, i.e.\ when $j_c(\theta)$ has a minimum
 at  $\theta=\pi/2$, a dip in $|M(H)|$ occurs in the region of weak
 fields. An additional dependence of $j_c$ on $|{\bf B}|$
 may then lead to a so-called fishtail (or peak) effect\cite{11}
 with a maximum of $|M|$ at some intermediate field.
 In Fig.~7 the zig-zag sweep of the applied field $H(t)$ has
 constant ramp rate, $|dH/dt| = 1$ in  units where
 $j_c(0)d =a =\mu_0 =E_0 =1$ in the used current--voltage law
 $E(J)= E_0 (J/J_c)^n$ with $n=51$, cf.\ Appendix B.
 When the creep exponent $n$ is lowered to a more realistic
 value $n=11$ measured e.g.\ in Ref.\ \onlinecite{26}, the
 amplitudes of the peak and dip slightly decrease, and the sharp
 edges of $M(H)$ at the largest and lowest $H_a$ are rounded, see
 also the figures in Refs.\ \onlinecite{19}d, \onlinecite{23,24}.

 \begin{figure}[F8]  
\epsfxsize= .90\hsize  \vskip 1.5\baselineskip
\centerline{ \epsffile{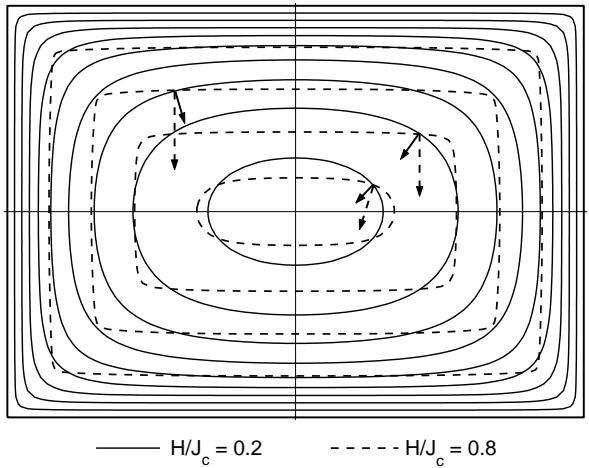}}
                       \vskip 0.5\baselineskip
 \caption{Stream lines of the sheet current in a
thin rectangular platelet with side ratio 1.4 and isoptropic
pinning. Bean model $J_c=$ const. Applied field $H/J_c =0.2$
(solid lines) and $H/J_c =0.8$ (dashed lines). Creep exponent
$n=21$. The arrows perpendicular to the stream lines and of
length proportional to the sheet current, indicate
the direction of the flux lines lying at different depths.
   }
\end{figure} 

 \section{Thin Rectangular Plates}

    The critical state of rectangular platelets is qualitatively
similar to that of the strip. In particular, for aspect ratios
$b/a >1.4$ (see Figs.~1, 8, 9) the profiles $J_y(x,0)$ and
$B_z(x,0)$ along the shorter axis are practically identical to the
profiles in long strips with $b/a \to \infty$. There are, however,
several qualitative differences as compared to the magnetic
behavior of long strips or circular disks:

a) In platelets with shape different from strips or circular
disks, the stream lines of the sheet current ${\bf J}(x,y)
=(J_x, J_y)$ do no longer coincide with the contour lines of
$B_z(x,y)$ and the penetrating flux fronts $\gamma_0$ (Fig.~1).
This new feature is described in detail in Ref.~4 for films with
elliptical shape.

b) In increasing applied field $H$, the direction of the sheet
current ${\bf J}(x,y)$ changes with $H$ at any given point
$(x,y)$ positioned inside the front $\gamma_0$ and away from the
symmetry axes $x=0$, $y=0$, or edges $x=\pm a$, $y=\pm b$. In the
isotropic case this rotation of ${\bf J}(x,y)$ comes to a halt
when the flux front passes through the point.

    Both features a) and b) occur only in perpendicular geometry.
In long rods with the same rectangular cross section in longitudinal
field, the $B_z$ contours, ${\bf J}$ stream lines, and flux fronts
are concentric rectangles, and inside the flux front both $B_z(x,y)$
and ${\bf J}(x,y)$ vanish. From the feature b) follows a further
interesting effect:

 \begin{figure}[F9]  
\epsfxsize= .98\hsize  \vskip 1.5\baselineskip
\centerline{ \epsffile{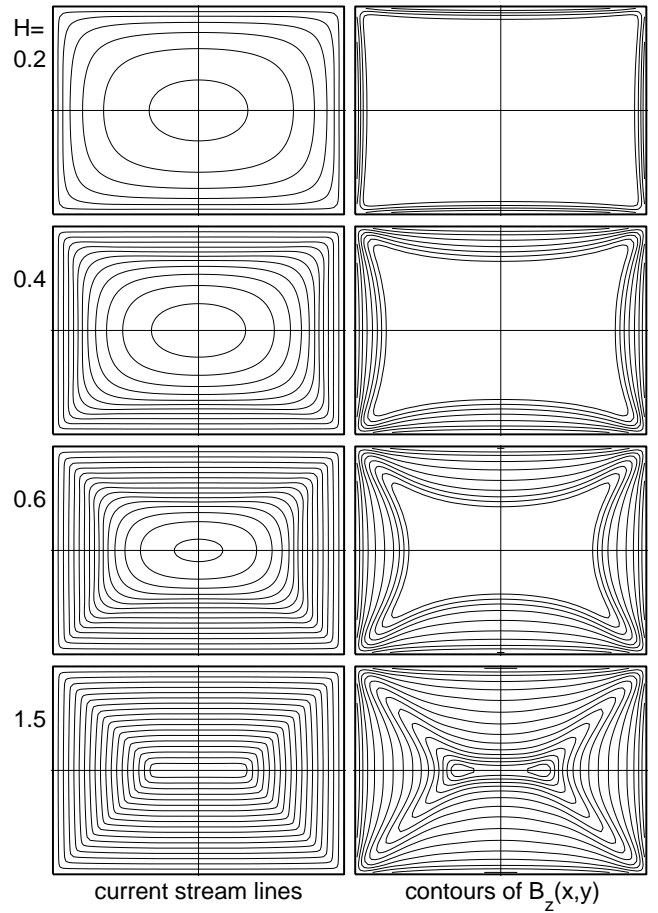}}
                       \vskip 0.5\baselineskip
 \caption{Stream lines of the sheet current ${\bf J}(x,y)$ (left)
 and contour lines of the perpendicular magnetic induction
 $B_z(x,y)$ (right) in a thin rectangular plate with out-of-plane
 anisotropic pinning and side ratio 1.4.  The anisotropy
 $j_c(\theta)$ is described by the model Eqs.~(17,18) with $p=2$
 and $q=4$. The  applied magnetic field is (from top to bottom)
 $H = 0.2$, 0.4, 0.6, 1.5 in units of $j_c(0) d$.
 Creep exponent $n=21$.
   } \end{figure}  

   In thin rectangles a {\it rotation} of the penetrating flux
lines may occur inside the front $\gamma_0$. In this region the
flux lines surround a thin flat field-free core and are nearly
parallel to the $x,y$ plane. However, at a given point $(x,y)$ the
flux lines lying at different $z$ can have different in-plane
orientations.  This rotation of flux lines is due
 to the rotation of the direction of the sheet current and
occurs in any isotropic or anisotropic thin superconductor with
shape different from a strip or circular disk. An example of this
rotation is shown in Fig.~8.

   The stream lines of the sheet current and the contours of
$B_z(x,y)$ in a thin rectangular plate with strong out-of-plane
anisotropy of pinning are depicted in Fig.~9 for increasing
applied field $H$. The stream lines are contours of the
function $g(x,y)$ introduced in Appendix B. In all contour plots
in Fig.~9 the same constant level spacing is used:
 $\Delta g = 1/20$ and $\Delta B_z = 1/10$ in units where
 $j_c(0)d = a = \mu_0 =1$. In our computation based on
Ref.~\onlinecite{7} we used $|dH/dt| =1$ and $E_0=1$
and $n=51$ in $E=E_0 (J/J_c)^n$. The shape of the depicted loops
is very similar to their shape in the isotropic plate,\cite{1,7}
i.e.\ different $J_c(B_z)$ dependences lead to almost the same
penetration pattern. This is so since the shape of these loops is
determined mainly by the specimen shape:
At small $H \ll J_c(0)$,
one has almost complete field expulsion and the stream lines are
thus independent of any material property, being rectangular near
the edges and circular near the center of the rectangle. After
the flux front has passed through a given point, $J$ abruptly tends
to nearly a constant, see Fig.~3 and Eq.~(17). The situation then
becomes similar to the isotropic case, and the stream lines practically
coincide with concentric rectangles with constant spacing
proportional to $1/J$.

  Since the current stream lines in the rectangular platelet do
not coincide with the lines $B_z=$const, and $J$ depends on $B_z$,
the penetration pattern in the anisotropic case could differ from
that obtained in the isotropic superconductor.\cite{1,7} However,
even a close look can hardly discover a qualitative difference in
the shapes of the contours in isotropic and strongly out-of-plane
anisotropic (Fig.~9) rectangular plates, in stark contrast to
the situation with in-plane-anisotropy.\cite{8} A minor difference
is that in the partly penetrated state ($H=0.4$ and $H=0.6$ in
Fig.~9)  the stream lines outside the flux front and near the axes
$x=0$ and $y=0$, are slightly convex towards the center, while for
isotropic Bean pinning the stream lines are straight there  (or
are slightly concave due to the finite creep exponent $n$). As
expected from the strip results, Fig.~3, with out-of-plane pinning
the density of the sheet-current stream-lines near the flux front
is larger (when $p>1$) or smaller (when $p<1$) as compared to the
case of isotropic pinning ($p=0$), cf.\ Eq.~(17). The same
statement is true for the contours of $B_z$ near the front.

  Figure 10 shows magnetization curves of thin rectangular
plates with side ratio 1.4 and various out-of-plane anisotropies.
These magnetization loops exhibit central peaks or dips caused by
the anisotropy and are very similar to those shown in Fig.~7
for strips.
Thus, one can conclude that the features
of the loops at $H\sim j_c d$ are specified mainly by the
anisotropy of flux-line pinning (i.e., by $p$ and $q$) rather than
by the shape of the sample. This property should be useful for
determining these pinning parameters from experiments.

 \section{Conclusions}

    In many papers thin superconductors are considered as 2D
 samples with a constant critical sheet current $J_c$ or with
 $J_c(B_z)=j_c(B_z)d$ if $j_c$ depends on the magnitude of the
 magnetic induction.  However, when an out-of-plane anisotropy of
 pinning exists, the current  density also depends on the angle
 $\theta$ between ${\bf B}$ (the flux-line direction) and the
 normal to the sample plane (our $z$ axis), and thus the critical
 state problem becomes three-dimensional.
 We have shown here that the general 3D critical state problem for
 thin superconductors of arbitrary shape and with arbitrary 3D
 anisotropy of pinning can be separated into a 1D problem, which
 treats the current density and induction across the thickness of
 the  sample, and the 2D problem of thin superconductors with a new
 induction dependence of the critical sheet current $J_c(B_z)$
 flowing in the plane of the superconductor. The new  $J_c(B_z)$
 dependence is determined by the out-of-plane anisotropy
 of the critical current density $j_c(\theta)$ and modifies the
 original $B_z$ dependence and in-plane anisotropy of $J_c$,
 if they exist. The resulting 2D thin film problem with given
 $J_c(B_z)$ can be solved by standard numerical and analytical
 methods.\cite{18,19,20,21,22,23,24} Our theory generalizes the
 approach of Ref.~\onlinecite{10} for a circular disk.

 \begin{figure}[F10]  
\epsfxsize= 0.90\hsize  \vskip 1.5\baselineskip
\centerline{ \epsffile{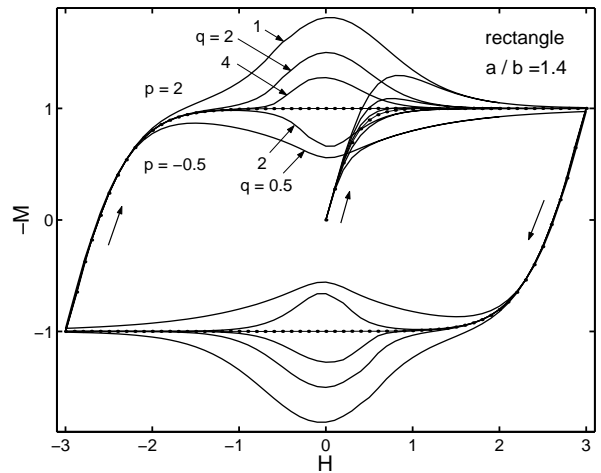}}
                       \vskip 0.5\baselineskip
 \caption{Magnetization curves of thin rectangular plates with
 anisotropic pinning, Eqs.~(17,18), side ratio $a/b=1.4$,
 and $n=51$.  As in Fig.~7, the dotted line
 is for isotropic pinning ($p=0$), the loops with a hump are for
 $p=2$, and the loops with a dip for $p=-0.5$. The applied field
 $H$ is in units of $j_c(0) d$ and $M$ is the magnetic moment
 in units of the saturation moment of the rectangle,
 $(2/3)j_c(0) d b^2(3a-b)$.
   } \end{figure}  

   Knowing the solutions of these 1D and 2D problems, the critical
 state in {\it three-dimensional} thin flat superconductors (Fig.~1)
 can be understood. Two qualitatively new features of the critical
 state occur in this general case as compared to the highly symmetric
 cases of isotropic disks or strips, which are commonly considered.
 First, when the shape of the sample differs from a strip or a disk
 then a {\it rotation} of flux lines (differently oriented in
 different planes $z=$ const) may occur in the region where
 $B_z=0$, i.e.\ where the flux lines surround a non-penetrated flat
 core and are nearly parallel to the sample plane since the sample
 is thin. Second, if both  out-of-plane and in-plane anisotropies
 of pinning exist, the flux lines even in {\it thin} superconductors
 are not only {\it curved} but also {\it twisted} in the region
 where $B_z\ne 0$, i.e.\ outside the front line $\gamma_0$ in
 Fig.\ 1. Twist means that the in-plane component of the direction
 of the flux lines at a given point $x,y$ changes with the depth $z$.

   Using the above approach, the critical states of strips and
 rectangular platelets with only out-of-plane anisotropy are
 analyzed. We consider a rather general analytical model (18) for
 this anisotropy, which can simulate both the intrinsic
 pinning by CuO planes and pinning by extended defects aligned
 with the $c$ axis in high-$T_c$ superconductors. We obtain the
 shape of the  flux front and the two-dimensional distributions
 of the magnetic induction and current density for the anisotropic
 strip. In the case of intrinsic pinning, the flux front has a
 wide flat part near the central plane, Fig.~5;
 the current density peaks near this flat front, and the induction
 profile is very steep there, see Fig.~3. In fact, a so-called current
 string\cite{27,28} occurs in this case, i.e., an additional current
 flows along this steep front as compared to the isotropic situation.
 In Refs.~\onlinecite{27,28} the string was considered in connection
 with a finite lower critical field $H_{c1}$.
 However, in the considered case the
 string is not due to a finite $H_{c1}$ but results from the
 anisotropy of pinning, and the jump of $B_z$ at the flux front
 considerably exceeds $H_{c1}$ since $H_{cr}=j_cd/2 \gg H_{c1}$
 was assumed above. This type of out-of-plane anisotropy
 leads to a central peak in the magnetization curve, Fig.~7.

   A different situation occurs when the $\theta$ dependence of
 $j_c$ has a minimum at $\theta=\pi/2$. In this case the flux front
 has the shape of a wedge with a rounded point that becomes sharper
 with increasing anisotropy. As can be seen in Figs.~3b,c the current
 density near the front line $\gamma_0$ now has no peak and no sharp
 step (as for isotropic pinning) but it decreases monotonically and
 has only a rounded step and an inflection point with vertical slope
 at this front line.
 The induction  $B_z$ now vanishes less steeply than in the
 isotropic case, and in the limit of large such anisotropy the
 induction profile near the front decreases almost linearly.
 The magnetization loop for this type of anisotropy has a central dip,
 see Fig.~7.

   All these features remain qualitatively the same when a
 smaller creep exponent $n$ is chosen in $E\propto J^n$, Eq.~(B6),
 or when the superconductor has a rectangular shape. In particular,
 with smaller $n$ the cusp in $J(x)$ remains sharp and the profile
 of $B_z(x)$ remains steep, see Fig.~3d with $n=11$. As shown in
 Figs.\ 7 and 10, the magnetization loops of thin rectangular
 plates with out-of-plane anisotropy exhibit the same central peak
 or dip as thin strips.

    For any given dependence  $j_c({\bf B})$,  the equations of
 this paper enable one to compute various characteristics of the
 critical state in thin superconducting samples of realistic shapes.
 With this theory at hand, experimental investigation of flux-density
 profiles, of the $H$ dependence of the penetrating flux front
 (e.g., by magneto-optics), and of magnetization loops, will yield
 information not only on the strengh but also anisotropy of
 flux-line pinning in superconductors.

 \acknowledgments

  G.P.M.~acknowledges the hospitality of the Max-Planck-Institute
f\"ur Metallforschung, Stuttgart.

 \appendix{ }   
 \section{electric field in anisotropic superconductors }

 In a uniform anisotropic superconductor the direction of the electric
 field ${\bf E}$ generated by vortex motion in general will not
 coincide with the direction of the current. To study this effect,
 we start with the situation when a current with the density $j$
 flows in the plane normal to the magnetic induction ${\bf B}$
 (below we call this plane the $N$ plane). The creep
 activation barrier $U[B, \varphi , j/j_0(B, \varphi)]$ is assumed to
 vanish at $j=j_0(B, \varphi)$ [this is the definition
 of the factor $j_0(B, \varphi)$]. Here the angle $\varphi$
 specifies the direction of the current in the $N$ plane,
 ${\bf j}\propto (\cos \varphi, \sin \varphi)$, and
 $U[B, \varphi, j/j_0(B, \varphi)]$ is the value of the barrier
 which a flux bundle has to overcome to hop in the direction of
 the Lorentz force, i.e., along the vector $[{\bf j}\times {\bf B}]$.
 At given temperature $T$, the expression
 $\exp\{-U[B, \zeta, j\cos(\zeta -\varphi)/j_0(B, \zeta)]/T\}$
 gives the probability for a flux-bundle to jump in the direction
 normal to any given in-plane vector $(\cos\zeta, \sin\zeta)$.
 Hence, the electric field ${\bf E}$ is obtained by averaging this
 expression over all angles $\zeta$ in the interval
 $\varphi -\pi/2\le \zeta \le \varphi +\pi/2$ (we disregard
 flux-bundle jumps against the Lorentz force).
    If the effective depth of the flux-pinning well
 $[\partial U(B, \varphi, x)/\partial x]_{x=1}$ considerably
 exceeds $T$, the direction of ${\bf E}$ in the $N$ plane,
 $(\cos \zeta_0,\,\sin \zeta_0)$, can be found as the position of the
 minimum of $U[B, \zeta, j\cos(\zeta-\varphi)/j_0(B, \zeta)]$
 with respect to variations of $\zeta$. Since, according to the
 definition of $j_0(B, \zeta)$, one has
 $\partial U(B, \zeta, 1)/\partial \zeta=0$, one obtains the
 following formula for $\zeta_0$ in the {\it critical state}:
    \[
    \tan(\zeta_0-\varphi)=-{j'_0(B, \zeta_0) \over j_0(B, \zeta_0)}\,,
    \]
 where $j'_0(B, \zeta)\equiv \partial j_0(B, \zeta)/\partial \zeta$.
 As for the critical current density $j_c(B, \varphi)$, we have
    \[
    j_c(B, \varphi)={j_0(B, \zeta_0) \over \cos(\zeta_0\!-\!\varphi)}=
    [j_0(B, \zeta_0)^2+j'_0(B, \zeta_0)^2]^{1/2} \,.
    \]
 It is also useful to express the angle $\zeta_0$ in terms of
 $j_c(B, \varphi)$, i.e., the quantity determined in experiments,
   \begin{equation} 
   \delta \equiv \zeta_0-\varphi =
   - \arctan \left\{ {j'_c(B, \varphi)
   \over j_c(B, \varphi)} \right\} \,,
   \end{equation}
 where $j_c'(B, \varphi)= \partial j_c(B, \varphi) /\partial\varphi$.
 The above formulas have been derived under the assumption that
 $\cos(\zeta-\varphi)/j_0(B_z, \zeta)$ has only one maximum in the
 interval $\varphi-\pi/2<\zeta<\varphi+\pi/2$. This situation does
 not always occur in twinned crystals. In such crystals the
 angle $\zeta_0$ may have a fixed value in some interval of
 angles $\varphi$, and thus
 $j_c(\varphi)=j_c(\zeta_0)/\cos(\zeta_0 -\varphi)$ inside the
 interval. However, note that Eq.~(A1) is still valid in this case.

   We now present the formulas for the case considered in Sec.\ II:
 critical currents flow in the $xy$ plane with
 ${\bf j}_c=(j_c\cos \varphi,\,j_c\sin \varphi)$,
 and a flux-line element has the direction
 $(\sin \theta \cos \psi,\,\sin \theta \sin \psi,\,\cos \theta)$.
 Since in the absence of flux cutting only the component of ${\bf j}_c$
 normal to ${\bf B}$ is essential, the critical
 current density in the $N$ plane, $j_c^{\perp}$, has the form
   \begin{equation} 
   j_c^{\perp}(B_z, B_t, \varphi\!-\!\psi, \psi)=
   n_{\perp}(\varphi)j_c(B_z, B_t, \varphi\!-\!\psi, \psi)\,,
   \end{equation}
 where $n_{\perp}(\varphi)=
 [1-\cos^2(\varphi-\psi)\sin^2\theta]^{1/2}$.
 It follows from Eq.~(A1) that the angle $\delta$ between
 ${\bf E}$ and ${\bf j}_c^{\perp}$ is given by the formula
   \begin{equation} 
   \tan\delta =-{\partial [\ln j_c^{\perp}(B_z, B_t, \varphi-\psi,
   \psi)] \over \partial \varphi}\cdot {n_{\perp}^2(\varphi) \over
   \cos\theta} \,.
   \end{equation}
 The unit vector ${\bf e}$ directed along ${\bf E}$ has the
 components:
   \begin{eqnarray} 
   e_x={1 \over n_{\perp}(\xi)}[\cos\xi-
   \sin^2\theta\cos\psi\cos(\xi-\psi)] \,, \nonumber \\
   e_y={1 \over n_{\perp}(\xi)}[\sin\xi-
   \sin^2\theta\sin\psi\cos(\xi-\psi)] \,,  \\
   e_z=-{1 \over n_{\perp}(\xi)} \sin\theta\cos\theta\cos(\xi-
   \psi) \,, \nonumber
   \end{eqnarray}
 where the angle $\xi$ is defined by the expression
   \begin{equation} 
   \tan(\xi-\psi) = \cos \theta {\tan(\varphi-\psi)+
   \cos\theta\tan\delta \over \cos\theta -
   \tan\delta\tan(\varphi-\psi)} \,.
   \end{equation}
 In other words, {\bf e} is the normalized projection of the vector
 $(\cos\xi,\,\sin\xi)$ onto the $N$ plane. This explains the meaning
 of the angle $\xi$.

 It is important to keep in mind one special case. If an
 in-plane anisotropy of flux-line pinning is absent, it follows
 from symmetry considerations that at fixed $B_z$, $B_t$, $\psi$
 the derivatives $\partial j_c^{\perp}/\partial \varphi$ and
 $\partial j_c/\partial \varphi$ are equal to zero when
 $\varphi -\psi =\pm \pi/2$. Thus, in this case one has $\delta =0$,
 $\xi =\varphi$, and the directions of ${\bf E}$, ${\bf j}_c$ and
 ${\bf j}_c^{\perp}$ coincide.

 \section{dynamic equations}  

   It is convenient to express the sheet current in the film
 through a scalar function $g(x,y) =g({\bf r})$ as
     \[
     {\bf J}({\bf r})=\nabla \times {\bf \hat z}g({\bf r})=
     -{\bf \hat z} \times\! \nabla g({\bf r})\,.
     \]
 This substitution guarantees that $\nabla \cdot {\bf J}=0$ and
 that the current flows along the specimen boundary $\Gamma$ if
 one puts $g({\bf r})={\rm const}=0$ there (the lines
 $g({\bf r})= {\rm const} $
 coincide with the current stream lines). Then, Eq.~(14) is
 transformed as follows,\cite{AB1}
    \begin{equation} 
   {B_z({\bf r}) \over \mu_0} =H +C({\bf r})g({\bf r}) -\!\int_S
   { g({\bf r'})- g({\bf r}) \over 4\pi R^3 }\, d^2  r' \,,
    \end{equation}
 with
     \begin{equation} 
    4 \pi C({\bf r})= \int_{\bar S} { d^2 r' \over R^3}
    = \int_0^{2 \pi} \!\!\!  {d\phi \over R(\phi)}\,.
    \end{equation}
 The integral in  Eq.~(B1) is taken in the Cauchy sense.
 In formulas (B2) the integration area $\bar S$ is the entire
 $x,y$ plane with exclusion of the sample area $S$;
 $R(\phi)\equiv |{\bf s}-{\bf r}|$;  ${\bf s}$
 defines the position of a point on the boundary $\Gamma$; and
 $\phi$ is the angle of the vector ${\bf s - r}$ relative to an
 arbitrary fixed direction. Eq.~(B1) admits inversion,
    \begin{equation} 
    g({\bf r})=\int_S\! K({\bf r}, {\bf r'}) \,
    [\, \mu_0^{-1} B_z({\bf r'}) -H \,] \, d^2 r' \,,
    \end{equation}
 where the kernel $K({\bf r}, {\bf r}_1)$ is found from
 the following equation,
    \begin{eqnarray} 
   K({\bf r, r}_1) ={1 \over C({\bf r})} \Big\{
   \delta({\bf r}-{\bf r}_1) ~~~    \nonumber \\
   + \!\int_S \! { K({\bf r',r}_1)- K({\bf r, r}_1) \over
   4\pi R^3 }\, d^2  r' \Big\} \,,
    \end{eqnarray}
 which can be solved by iteration starting with $K({\bf r, r}_1)=0$.

   We now differentiate Eq.~(B3) with respect to time $t$, take
 into account the Maxwell equation,
 $\partial B_z/\partial t=-{\bf \hat z}\cdot [\nabla \times {\bf E}]$,
 and the current-voltage law,\cite{AB2}
 \[
 {\bf E}={\bf E}({\bf J}, B_z) \,,
 \]
 which expresses the electric field ${\bf E}$ through the sheet
 current ${\bf J}$.  Then, we arrive at an equation
 for $g({\bf r})$,
    \begin{eqnarray} 
    {\partial g({\bf r},t) \over \partial t} =
    \int_S K({\bf r}, {\bf r'}) ~~~~~~~~~~~~~~~~~~~ \nonumber \\
    \left\{ -[ {\bf \hat z}\times \nabla ] \cdot
    {\bf E}[-{\bf \hat z} \times\! \nabla g({\bf r'},t),
    B_z({\bf r'}) ]
    -{\partial H(t) \over \partial t} \right\} d^2 r' .
    \end{eqnarray}
 Equations (B1) and (B5) allow one to find $B_z({\bf r},t)$,
 $g({\bf r},t)$ [and thus ${\bf J}({\bf r},t)$] if the
 prehistory $H(t)$ and the current-voltage law  are given.

   The static critical state can be calculated by a dynamic
 approach using any model law ${\bf E}$ which has a sharp bend
 at $J = J_c$. The specific form of this dependence is irrelevant;
 e.g.\  we may use the power law
     \begin{equation} 
     {\bf E} ={\bf E}_0\left ({|{\bf J}| \over J_c}\right )^n
     \end{equation}
 and take the limit $n\gg 1$.   That the static critical state is
 indeed reached in the limit $n\to \infty$ was shown rigorously in
 Ref.\ \onlinecite{AB3}.
 The function $J(B_z,\varphi_0)$ resulting from Eq.~(12) or from
 the appropriate generalization of this equation
 should be used as the critical sheet current $J_c$ in Eq.~(B6)
 for currents with direction $(\cos \varphi_0, \sin \varphi_0)$.
 The direction of the constant vector ${\bf E}_0$
 in general does not coincide with ${\bf J}$ if there is
 anisotropy of flux-line pinning in the $xy$ plane (see Appendix A).
 At given $\varphi_0$, the angle $\xi_0$ defining the direction of
 ${\bf E}\propto (\cos\xi_0, \sin\xi_0)$ is determined by the
 formula
   \begin{equation} 
   \tan(\xi_0-\varphi_0)=-{\partial [\ln J(B_z, \varphi_0)]
   \over \partial \varphi_0}\ .
   \end{equation}

 Thus, to describe the critical state of an
 infinitely thin superconductor, it is sufficient to solve
 the dynamical Eqs.~(B1,B5,B6) with the initial condition
 $g(x,y)|_{t=0} =0$. This dynamical approach appears to be
 more convenient than solving the static Eqs.~(13,14) with
 an initially unknown flux-front $\gamma_0$.


 \end{multicols}

\begin{thebibliography}{1}

\bibitem{1} E. H. Brandt, Rep. Prog. Phys. {\bf 58}, 1465 (1995).

\bibitem{2} P.N. Mikheenko and Yu.E. Kuzolev, Physica C {\bf 204},
            229 (1994).

\bibitem{3} E.H. Brandt, M.V. Indenbom and A. Forkl, Europhys.
            Lett. {\bf 22}, 735 (1993); E.H. Brandt and M.V. Indenbom,
            Phys. Rev. B {\bf 48}, 12893 (1993); E. Zeldov, J.R. Clem,
            M. McElfresh and M. Darwin, Phys. Rev. B {\bf 49}, 9802,
            (1994).

\bibitem{4} G.P.Mikitik and E.H. Brandt, Phys. Rev. B {\bf 60},
            592 (1999).

\bibitem{5} E.H. Brandt, Phys. Rev. Lett. {\bf 74}, 3025 (1995).

\bibitem{6} L.~Prigozhin, J. Comp. Phys. {\bf 144}, 180 (1998).

\bibitem{7} E.H. Brandt, Phys. Rev. B {\bf 52}, 15442 (1995).

\bibitem{8} Th. Schuster, H. Kuhn, E.H. Brandt, and S. Klaum\"unzer,
            Phys. Rev. B {\bf 56}, 3413 (1997).

\bibitem{9} M.~Tachiki and S.~Takahashi, Solid State Commun.
            {\bf 70}, 2991 (1991).

\bibitem{10} I.M. Babich and G.P. Mikitik, Phys. Rev. B {\bf 54},
            6576 (1996).

\bibitem{11} I.M. Babich and G.P. Mikitik, Pis'ma Zh. Eksp. Teor.
             Fiz. {\bf 64}, 538 (1996) [JETP Lett. {\bf 64}, 586 (1996)];
             I.M. Babich and G.P. Mikitik, Phys. Rev. B {\bf 58},
             14207 (1998).

\bibitem{12} E.~Zeldov, A.~I.~Larkin, V.~B.~Geshkenbein,
           M.~Konczykowski, D.~Majer, B.~Khaykovich, V.~M.~Vinokur,
           and H.~Strikhman, \prl{\bf 73}, 1428 (1994);
       M.\ Benkraouda and J.\ R.\ Clem, \prb{\bf 53}, 5716 (1996);
       R.\ Labusch and T.\ B.\ Doyle, Physica C {\bf 290}, 143
       (1997); T.\ B.\ Doyle, R.\ Labusch, and R.\ A.\ Doyle,
       Physica C {\bf 290}, 148 (1997);
       E.~H.~Brandt, \prb{\bf 59}, 3369 (1999);
       ibid, \prb{\bf 60}, 11939 (1999).

\bibitem{13} To be able to
 consider the superconductor as a uniform medium, the
 characteristic distance between the twin boundaries, $l_{tw}$,
 should be less than the scale over which the currents and the fields
 are averaged, i.e., at least $l_{tw}\ll d$.

\bibitem{14} A.~M.~Campbell and J.~E.~Evetts, Adv.\ Phys.\ {\bf 72},
             199 (1972).

\bibitem{15} J.~R.~Clem, \prb{\bf 26}, 2463 (1982);
             J.~R.~Clem and A.~ Perez-Gonzalez, \prb{\bf 30}, 5041 (1984);
         A.~Perez-Gonzalez and J.~R.~Clem, \prb{\bf 31}, 7048 (1985).

\bibitem{16} E.~H.~Brandt, Phys.\ Lett.\ {\bf 79A}, 207 (1980);
             E.~H.~Brandt, J.\ Low Temp.\ Physics {\bf39}, 41 (1980);
         E.~H.~Brandt, J.\ Low Temp.\ Physics {\bf44}, 33, 59 (1981).

\bibitem{17} L.~W.~Conner, A.~P.~Malozemoff, and I.~A.~Campbell,
            Phys. Rev. B {\bf 44}, 403 (1991).

\bibitem{18} E.~H.~Brandt, \prb{\bf 49}, 9024 (1994).

\bibitem{19} E.~H.~Brandt, \prb{\bf 50}, 4034 (1994);
         Th.~Schuster et al., \prl{\bf 73}, 1424 (1994);
         E.~H.~Brandt, Physica C {\bf 235-240}, 2939 (1994);
         \prb{\bf 55}, 14513 (1997).

\bibitem{20} G.~P.~Mikitik and E.~H.~Brandt, Phys. Rev. B (submitted).

\bibitem{21} J.~McDonald and J.~R.~Clem, \prb{\bf 53}, 8643 (1996).

\bibitem{22} D.~V.~Shantsev, Y.~M.~Galperin, and T.~H.~Johansen,
             \prb{\bf 60}, 13112 (1999).

\bibitem{23} E.~H.~Brandt, Phys.\ Rev.\ B {\bf 54}, 4246 (1996).

\bibitem{24} E.~H.~Brandt, Phys.\ Rev.\ B {\bf 58}, 6506 (1998).

\bibitem{25} See, e.g., D.~V.~Shantsev, M.~R.~Koblischka,
        Y.~M.~Galperin, T.~H.~Johansen, L.~Pust, and M.~Jirsa,
        \prl{\bf 82}, 2947 (1999) and references therein.

\bibitem{26} B.~J.~J\"onsson-{\AA}kerman, K.~V.~Rao, and
        E.~H.~Brandt, Phys.\ Rev.\ B {\bf 60}, 14913 (1999).

\bibitem{27} M.~V.~Indenbom, Th.~Schuster, H.~Kuhn, H.~Kronm\"uller,
        T.~W.~Li, A.~A.~Menovsky, \prb{\bf 51}, 15484 (1995).

\bibitem{28} E.~H.~Brandt, \prb{\bf 59}, 3369 (1999).

\bibitem{AB1} E.~H.~Brandt, Phys.\ Rev.\ B {\bf 46}, 8628 (1992).

\bibitem{AB2} Note that we replace the current density ${\bf j}$ by
 the sheet current ${\bf J}$ in the original current-voltage law
 ${\bf E}({\bf j},B_z)$. This is justified by the following
 considerations. First, in this paper the dynamical approach
 is only a useful trick which allows us to obtain the
 solution of the critical state equations.
 Second, in the leading order in $d/L$, the electric
 field ${\bf E}(x,y,z)$ is a constant across the thickness of
 the sample in the region outside the flux front [this estimate is
 also confirmed by the results of computations for strips and
 disks of a finite thickness.\cite{23,24}
 Thus, if one replaces $j/j_c$ by $J/J_c$ in the current-voltage
 law, this is really a good approximation in solving the dynamic
 equations for thin superconductors.

\bibitem{AB3} J.~W.~Barrett and L.~Prigozhin, Nonlinear Analysis.
 Theory, Methods and Applications (in print) (2000).

\end{thebibliography}
 \end{document}